# Granular materials flow like complex fluids

Binquan Kou[1], Yixin Cao[1], Jindong Li[1], Chengjie Xia[1], Zhifeng Li[1], Haipeng Dong[4], Ang Zhang[4], Jie Zhang[1,5], Walter Kob[2]*, and Yujie Wang[1,3,6]*

[1]School of Physics and Astronomy, Shanghai Jiao Tong University, 800 Dong Chuan Road, Shanghai 200240, China
[2]Laboratoire Charles Coulomb, University of Montpellier and CNRS, UMR 5221, 34095 Montpellier, France
[3]Materials Genome Initiative Center, Shanghai Jiao Tong University, 800 Dong Chuan Road, Shanghai 200240, China
[4]Department of Radiology, Ruijin Hospital, Shanghai Jiao Tong University School of Medicine, Shanghai 200025, China
[5] Institute of Natural Sciences, Shanghai Jiao Tong University, Shanghai 200240, China
[6] Collaborative Innovation Center of Advanced Microstructures, Nanjing University, Nanjing, 210093, China

**Granular materials such as sand, powders, foams etc. are ubiquitous in our daily life, as well as in industrial and geotechnical applications[1-4]. Although these disordered systems form stable structures if unperturbed, in practice they do relax because of the presence of unavoidable external influences such as tapping or shear. Often it is tacitly assumed that for granular systems this relaxation dynamics is similar to the one of thermal glass-formers[3,5], but in fact experimental difficulties have so far prevented to determine the dynamic properties of three dimensional granular systems on the particle level. This lack of experimental data, combined with the fact that in these systems the motion of the particles involves friction, makes it very challenging to come up with an accurate description of their relaxation dynamics. Here we use X-ray tomography to determine the microscopic relaxation dynamics of hard granular ellipsoids that are subject to an oscillatory shear. We find that the distribution function of the particle displacement can be described by a Gumbel law[6] with a shape parameter that is independent of time and the strain amplitude $\gamma$. Despite this universality, the mean squared displacement of a tagged particle shows power-laws as a function of time with an exponent that depends on $\gamma$ and the time interval**

**considered. We argue that these results are directly related to the existence of the microscopic relaxation mechanisms that involve friction and memory effects. These observations demonstrate that on the particle level the dynamical behavior of granular systems is qualitatively different from the one of thermal glass-formers and instead more similar to the one of complex fluids. Thus we conclude that granular materials can relax even when the driving is weak, an insight which impacts our understanding of the nature of granular solids.**

Static and driven granular systems behave in many ways like thermal glassy systems[3]. However, they also show phenomena like arching and force chains[2] as well as shear thickening and avalanches which demonstrates that their properties are also affected by unique many-particle effects. Although there are a multitude of studies on the macroscopic properties of granular systems, only few investigations have probed in three dimensions (3D) their structure and dynamics on the level of the particles because of the experimental difficulty to determine the position and orientation of the particles[7-9]. This information is, however, crucial to obtain a fundamental understanding of the properties of these systems on a macroscopic level[2]. Dynamical information on the system can be obtained by cyclic shear experiments/simulations which allow to obtain direct insight into the compaction and aging dynamics[10,11], memory effect[12], dynamic heterogeneity[13], plastic deformation[14], and avalanches[15], as well as the role of friction[11,16,17]. Our X-ray tomography study shows that this dynamics gives rise to a universal distribution of the particle displacements, described well by a Gumbel law[6], as well as marked memory effects. Surprisingly, we find that the cage effect, the mechanism that is responsible for the slowing down of the dynamics in glass-forming system[5], is absent, thus indicating a marked difference in the microscopic relaxation dynamics of thermal glass formers and granular systems.

The particles we study are hard plastic prolate ellipsoids (minor axis $2b = 12.7$mm) with an aspect ratio of 1.5. The particles are poured into a rectangular box with side walls that can be tilted to impose a cyclic shear on the system and a heavy plate on the top of the particles (see SI for details). We have used four strain amplitudes, $\gamma$ =0.07, 0.10, 0.19, and 0.26, and a strain rate of around $1.7 \times 10^{-2} s^{-1}$, i.e. the inertial number is about $1.4 \times 10^{-4}$ indicating that the experiment can be considered as quasi-static[18]. Before taking any measurements, we have made hundreds of cycles to bring the system into a stationary state (see SI for details). After each cycle we do a tomography scan and measure the position and orientations of all the particles[19] (Fig 1a. The movies included in the SI show the dynamics of the particles). In the following we will measure time *t* in units of complete cycles and length in units of b.

The translational mean squared displacement (TMSD), $\langle d^2(t) \rangle = \langle |\mathbf{r}_j(t) - \mathbf{r}_j(0)|^2 \rangle$ of the particles are shown in Fig. 1b. We see that the TMSD has no plateau at intermediate times even if the driving is weak which demonstrates that the cage effect observed in thermal glassy systems[5] is absent. This is also confirmed by the intermediate scattering function which does not show any sign of a two step relaxation (see SI). For $\gamma$ =0.07, $\langle d^2(t) \rangle$ is proportional to *t* for all times accessed, indicating normal diffusive behavior. If $\gamma$ is increased to 0.26, the *t*-dependence of $\langle d^2(t) \rangle$ is a power-law with an exponent around 0.75, i.e. the diffusion is anomalous. This result is somewhat surprising, since naively one could expect that a stronger driving induces more noise to the system and hence leads to a diffusive motion. This is obviously not the case and below we will discuss the origin of this behavior. For $\gamma$ =0.10 and 0.19 we find for short times a power-law with an exponent close to 1.0, i.e. normal diffusion, but surprisingly at $t \approx 100$ and 10, respectively, this *t*-dependence crosses over to the power-law with exponent close to 0.75, i.e. the same exponent we found for $\gamma$ =0.26. Possibilities to explain the presence of such an anomalous diffusion are to assume that the underlying configuration space has a fractal nature[20], that the dynamics shows aspects of a Levy flight[21], or that the motion of the particles is collective[22-25]. In the following we will show, however, that here the anomalous diffusion can be rationalized by simple memory effect in the dynamics.

Also the rotational dynamics at short times shows for $\gamma=0.26$ anomalous diffusion, Fig. 1c. However, at around $t \approx 40$ it crosses over to a linear $t$-dependence indicating that for this strain amplitude the particles start the rotational diffusion well before the translational diffusion sets in. For $\gamma=0.19$ the anomalous diffusion in the rotational dynamics is less pronounced (i.e. the exponent is larger) and for $\gamma \leq 0.1$ we find normal diffusion for all times. From these observations we thus can conclude that also the rotational part of phase space and the manner it is explored depends significantly on $\gamma$.

More detailed information on the relaxation dynamics can be obtained from the distribution function for the displacements, i.e. the van Hove function $G_s(d,t) = \frac{1}{N}\sum_{j=1}^{N}\langle \delta(d - |\mathbf{r}_j(t) - \mathbf{r}_j(0)|)\rangle$ [5], shown in Fig. 2a. Although these curves suggest that $G_s(d,t)$ has a similar shape as the one found in thermal glass-forming systems[5,26], a plot of the distribution of the displacements along a given axis demonstrates that it is not given by a Gaussian, but can instead be fitted very well by a Gumbel law[6],

$$f(d_y) = A(\lambda)\exp\left(-\frac{|d_y|}{\lambda} - \exp\left(-\frac{|d_y|}{\lambda}\right)\right),$$

Fig. 2b. (See SI on an alternative q-Gaussian fit). Here $\lambda$ is a length scale the nature of which we discuss in the SI. There we also show that this Gumbel law can be interpreted as the consequence of the interplay of two relaxation mechanisms: Diffusion on small length scales that is induced by the roughness of the particles and by friction, and irreversible relaxation events on somewhat larger scales. Since this functional form seems to be also compatible with observations from other granular systems, see e.g. Ref. 15, we can conclude that our findings are not just a particularity of the present system, but more general. Most remarkable is the fact that the shape of the distribution is independent of $t$ and $\gamma$ (see Fig. 2b), as well as the direction of the displacement, which suggests that the mechanism leading to this distribution is very general and in the SI we give arguments why this is the case.

That the shape of $G_s(d,t)$ is independent of $\gamma$ and $t$ indicates that the anomalous diffusion is not related to the details of this distribution and more generally that a one-time quantity is not able to explain the anomalous TMSD. Therefore we look at a generalization of this dynamic observable and define a two-time correlation function $G_4(d_1,d_2,t_w) = \left\langle \delta\left[d_1 - (r_y(t_w) - r_y(0))\right] \delta\left[d_2 - (r_y(2t_w) - r_y(t_w))\right] \right\rangle$. This function gives thus the probability that if a particle has moved by $d_1$ in the time interval $[0,t_w]$, it will move by $d_2$ in the interval $[t_w, 2t_w]$. For a Markovian process the joint probability will factorize and thus it can be used to test for the presence of non-Markovian behavior, such as memory effects[27]. The scatter plots for this joint distribution, Fig. 3, shows that for $\gamma = 0.10$, Fig. 3a, there is a reflection symmetry around the horizontal axis indicating that for any given value of $d_1$, the probability to find a positive $d_2$ is the same as for a negative $d_2$ (The $t_w$-dependence of these plots will be discussed in Fig. 4.). By dividing the $d_1$ axis into three ranges (see Fig. 3a) we can compute for these values of $d_1$ the conditional probability that a particle moves a distance $d_2$. The resulting three distributions are shown in Fig. 3b and we see that they are independent of the range considered, which demonstrates that the joint distribution can indeed be factorized. Figures 3c and 3d show the same quantities for $\gamma = 0.26$. We recognize that now the joint distribution is aligned with the negative diagonal, thus indicating the presence of a memory effect, *i.e.* a particle that has moved in time $t_w$ by an amount $d_1$ is likely to move in the time interval $[t_w, 2t_w]$ by an amount $-d_1$[13,27]. This can be seen clearly in the coarse-grained distributions (Fig. 3d) which are now no longer symmetric but become increasingly skewed with increasing $d_1$. Since the motion during the interval $[t_w, 2t_w]$ is backwards, the particles advance a bit slower than can be expected from $G_s(d_1,t)$ and thus this gives rise to the anomalous diffusion. To quantify how this memory effect changes with $\gamma$ and $t_w$ we divide the $d_1$ axis into intervals of width $0.5\langle d_y^2 \rangle^{1/2}$ and compute for each interval the average value of $d_2$. If this average is zero the joint probability factorizes whereas a non-zero value signals a non-Markovian behavior. Figure 4a shows this mean value for different times and we see that for $\gamma=0.10$ and short times the memory effect is very small, i.e. the process is Markovian. In contrast to this we find that at large times, and for $\gamma=0.26$ for all times, the memory is quite pronounced thus confirming the non-Markovian character of the dynamics when the particles start to explore a region that becomes comparable with their size.

In order to study the duration of this memory effect we have calculated the mean value of $d_2$ (considering only $d_1>0$ since there is a point symmetry with respect to the origin) and in Fig. 4b we show its $t_w$-dependence. The curve for $\gamma=0.10$ demonstrates that the memory effect starts to become noticeable at $t_w \approx 100$, i.e. when the TMSD makes a cross-over from the simple diffusion motion to the anomalous diffusion (see Fig. 1b), indicating that these two phenomena are indeed related. The curve for $\gamma=0.26$ shows that the memory effect last significantly longer than 200 cycles, *i.e.* a time span on which the particles have undergone a translational motion that is a significant fraction of their size (Fig. 1a).

Our results demonstrate that the relaxation dynamics of cycled granular systems is qualitatively very different from the one of thermal glass-formers in that it does not show the cage effect. Instead our observations indicate that the presence of tiny relaxation events as well as friction during the cycle gives rise to a distribution of the displacements that can be described well by a Gumbel law. These results and the fact that for intermediate sized displacements of the particles one sees marked memory effects suggest that such granular materials are similar to so-called "complex fluids", i.e. liquids that have a relaxation dynamics that is non-Debye and often involve memory effects[28]. Since the mechanism leading to the observed relaxation dynamics is very general we conclude that in general granular systems will usually relax even if they are just slightly perturbed despite the fact that if not driven they undergo jamming transition and become solids. Thus the perturbation makes that a granular system behaves like a continuously evolving solid which steadily explores the mechanically stable phase space allowed by static friction. This insight should thus allow to connect the relaxation dynamics on the particle level to mesoscopic phenomena like secondary creep and non-local rheology[29] and thus to lay the groundwork to rationalize the macroscopic behavior of granular systems in terms of the microscopic properties.

## Methods

**Experimental Details**

The granular particles are prolate ellipsoids with an aspect ratio of 1.5, a size polydispersity around 0.9%, and are made of polyvinyl chloride (PVC) by injection molding. The major and minor axes are 2a=19mm and 2b=12.67mm, respectively. During the course of the experiment we have found no evidence of crystallization (see the static structure factor $S(q)$ in SI). This good glass-forming ability is likely related to the fact that ellipsoids with this aspect ratio can be packed to an amorphous phase that is relatively dense[30]. There are about 4100 ellipsoids in the shear cell and the initial steady state of the system was prepared by first making hundreds of shear cycles (see SI for details on the cell and the preparation of the system). The three-dimensional structural information (position and orientation of all the particles) was acquired by a medical CT scanner (SOMATOM Perspective, Siemens, Germany) with a spatial resolution of 0.6mm. One entire scan took about 10s. We took a tomography scan after each cycle for the first 10 cycles and then a scan every 5 cycles. For $\gamma=0.10$ we made a second experiment in which we scanned only every 50 cycles allowing thus to reach larger times.

We followed similar imaging processing steps as in previous studies[19]. Through a marker-based watershed image segmentation technique, we extracted the center position $r(t)$ and orientation $e(t)$ of each ellipsoid. By making two consecutive tomography scans of the same static packing and analyzing them independently, the precision of the extracted center position $r(t)$ and orientation $e(t)$ can be estimated to be $5.3\times10^{-3}$b and $8.4\times10^{-3}$rad, respectively. To minimize the influence of boundary effects on our results we have considered only those particles that have a distance of at least 5b from the cell boundaries (see SI for details). This condition makes that in practice we used about 1300 particles for the analysis.

## References

1       Jaeger, H. M., Nagel, S. R. & Behringer, R. P. Granular solids, liquids, and gases. *Rev. Mod. Phys.* **68**, 1259-1273 (1996).

2       Duran, J. *Sands, powders, and grains: an introduction to the physics of granular materials*.  (Springer Science & Business Media, 2012).

3       Coniglio, A., Fierro, A., Herrmann, H. J. & Nicodemi, M. *Unifying concepts in granular media and glasses*.  (Elsevier, 2004).

4       Anthony, J. L. & Marone, C. Influence of particle characteristics on granular friction. *J. Geophys. Res.* **110**, 1-14 (2005).

5       Binder, K. & Kob, W. *Glassy materials and disordered solids: An introduction to their statistical mechanics*.  (World Scientific, 2011).

6       Bramwell, S. T. The distribution of spatially averaged critical properties. *Nat. Phys.* **5**, 444-447 (2009).

7       Mueth, D. M. *et al.* Signatures of granular microstructure in dense shear flows. *Nature* **406**, 385-389 (2000).

8       Dijksman, J. A., Rietz, F., Lorincz, K. A., Van Hecke, M. & Losert, W. Refractive index matched scanning of dense granular materials. *Rev. Sci. Instrum.* **83**, 011301 (2012).

9       Panaitescu, A., Reddy, K. A. & Kudrolli, A. Nucleation and Crystal Growth in Sheared Granular Sphere Packings. *Phys. Rev. Lett.* **108**, 108001 (2012).

10      Pouliquen, O., Belzons, M. & Nicolas, M. Fluctuating Particle Motion during Shear Induced Granular Compaction. *Phys. Rev. Lett.* **91**, 014301 (2003).

11      Ren, J., Dijksman, J. A. & Behringer, R. P. Reynolds pressure and relaxation in a sheared granular system. *Phys. Rev. Lett.* **110**, 018302 (2013).

12      Paulsen, J. D., Keim, N. C. & Nagel, S. R. Multiple transient memories in experiments on sheared non-Brownian suspensions. *Phys. Rev. Lett.* **113**, 068301 (2014).

13      Dauchot, O., Marty, G. & Biroli, G. Dynamical Heterogeneity Close to the Jamming Transition in a Sheared Granular Material. *Phys. Rev. Lett.* **95**, 265701 (2005).

14      Slotterback, S. *et al.* Onset of irreversibility in cyclic shear of granular packings. *Phys. Rev. E* **85**, 021309 (2012).

15   Radjai, F. & Roux, S. Turbulentlike fluctuations in quasistatic flow of granular media. *Phys. Rev. Lett.* **89**, 064302 (2002).

16   Bi, D., Zhang, J., Chakraborty, B. & Behringer, R. P. Jamming by shear. *Nature* **480**, 355-358 (2011).

17   Royer, J. R. & Chaikin, P. M. Precisely cyclic sand: self-organization of periodically sheared frictional grains. *Proc. Natl Acad. Sci. USA* **112**, 49-53 (2015).

18   GDR MiDi. On dense granular flows. *Eur. Phys. J. E* **14**, 341-365 (2004).

19   Xia, C. *et al.* The structural origin of the hard-sphere glass transition in granular packing. *Nat. Commun.* **6**, 8409 (2015).

20   Mailman, M., Harrington, M., Girvan, M. & Losert, W. Consequences of Anomalous Diffusion in Disordered Systems under Cyclic Forcing. *Phys. Rev. Lett.* **112**, 228001 (2014).

21   Bouchaud, J. P. & Georges, A. Anomalous diffusion in disordered media: Statistical mechanisms, models and physical applications. *Phys. Rep.* **195**, 127-293 (1990).

22   Singh, A., Magnanimo, V., Saitoh, K. & Luding, S. The role of gravity or pressure and contact stiffness in granular rheology. *New J. Phys.* **17**, 043028 (2015).

23   Kumar, N. & Luding, S. Memory of jamming-multiscale models for soft and granular matter. *Granul Matter* **18**, 58 (2016).

24   Henann, D. L. & Kamrin, K. Continuum modeling of secondary rheology in dense granular materials. *Phys. Rev. Lett.* **113**, 178001 (2014).

25   Caballero-Robledo, G. A., Goldenberg, C. & Clement, E. Local dynamics and synchronization in a granular glass. *Granul Matter* **14**, 239-245 (2012).

26   Kob, W. & Andersen, H. C. Testing mode-coupling theory for a supercooled binary Lennard-Jones mixture I: The van Hove correlation function. *Phys. Rev. E* **51**, 4626-4641 (1995).

27   Doliwa, B. & Heuer, A. The origin of anomalous diffusion and non-Gaussian effects for hard spheres: Analysis of three-time correlations. *J. Phys. Condens. Matter* **11**, A277-A283 (1999).

28   Larson, R. G. *The structure and rheology of complex fluids*. (Oxford University Press, 1999).

29   Kamrin, K. & Koval, G. Nonlocal constitutive relation for steady granular flow. *Phys. Rev. Lett.* **108**, 178301 (2012).

30   Donev, A. *et al.* Improving the density of jammed disordered packings using ellipsoids. *Science* **303**, 990-993 (2004).

*Supplementary information* is available in the online version of the paper.

*Acknowledgments:* Some of the preliminary experiments were carried out at BL13W1 beamline of Shanghai Synchrotron Radiation Facility. The work is supported by the National Natural Science Foundation of China (No. 11175121, 11675110 and U1432111), Specialized Research Fund for the Doctoral Program of Higher Education of China (Grant No. 20110073120073) and ANR-15-CE30-0003-02.. W.K. is member of the Institut Universitaire de France.

*Author contributions:* Y.W. and W.K. designed the research. B.K., Y.C., J.L., C.X., Z.L., H.D., A.Z., J.Z. and Y.W. performed the experiment. B.K., W.K. and Y.W. analyzed the data and wrote the paper.

*Author Information:* Reprints and permissions information is available at www.nature.com/reprints. The authors declare no competing financial interests. Readers are welcome to comment on the online version of the paper. Correspondence and requests for materials should be addressed to: Y.W. (yujiewang@sjtu.edu.cn) or W.K. (walter.kob@ umontpellier.fr)

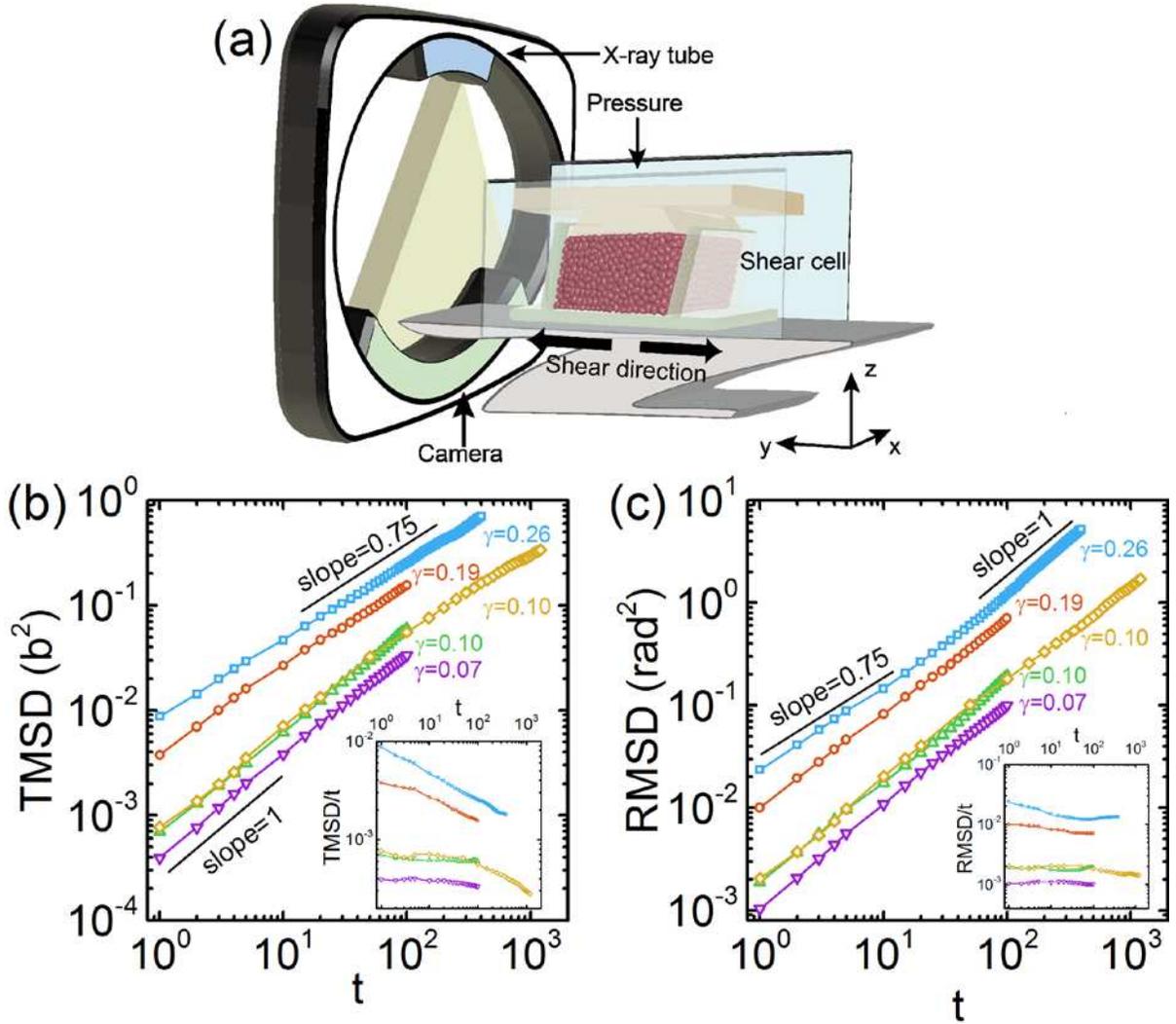

**Figure 1| Experimental setup and time dependence of the translational and rotational mean squared displacements for different strain amplitude $\gamma$ to show the presence of anomalous diffusion.**

a) Schematic picture of the experimental setup showing the CT scanner and the box containing the probed granular media. Also indicated are the directions of the coordinate axis used in the present study. b) The translational mean squared displacement (TMSD) for $\gamma$ =0.07, 0.10, 0.19, and 0.26 shows a power-law behavior with an exponent that depends on the time interval considered. For $\gamma$ =0.10 we have two set of curves from two independent measurements: The first one (green curve) is probing the short time dynamics. The orange curve uses a larger time grid and thus shows the TMSD at long times. From the fact that the two curves superimpose very well at intermediate times we can infer that the precision of the data is high. Inset: TMSD/$t$ as a function of $t$ to demonstrate that for small $\gamma$ and short times the TMSD is indeed linear in time whereas for large times and large $\gamma$ the dynamics is subdiffusive. Also note the kink in the curves for $\gamma=0.19$ (and $\gamma=0.10$) at around $t=10$ (around $t=100$) that indicates that the relaxation dynamics of the system changes from normal diffusion to anomalous diffusion. The anomalous diffusion sets in once the particles start to move distances that are on the order of their size. This behavior can thus be caused by the shear during one cycle with large strain, see $\gamma=0.26$, or by the diffusive motion after many cycles ($\gamma \leq 0.19$). Note that for times larger than the ones accessed in these experiments the TMSD will become linear in time for all values of $\gamma$, i.e. the particle will ultimately become diffusive. c) Rotational mean squared displacement (RMSD) for $\gamma=0.07$, 0.10, 0.19, and 0.26 showing that at short times the exponent of the power-law depends on $\gamma$. Inset: RMSD/$t$ as a function of $t$ showing that at large times the RMSD is linear in time for all shear amplitudes.

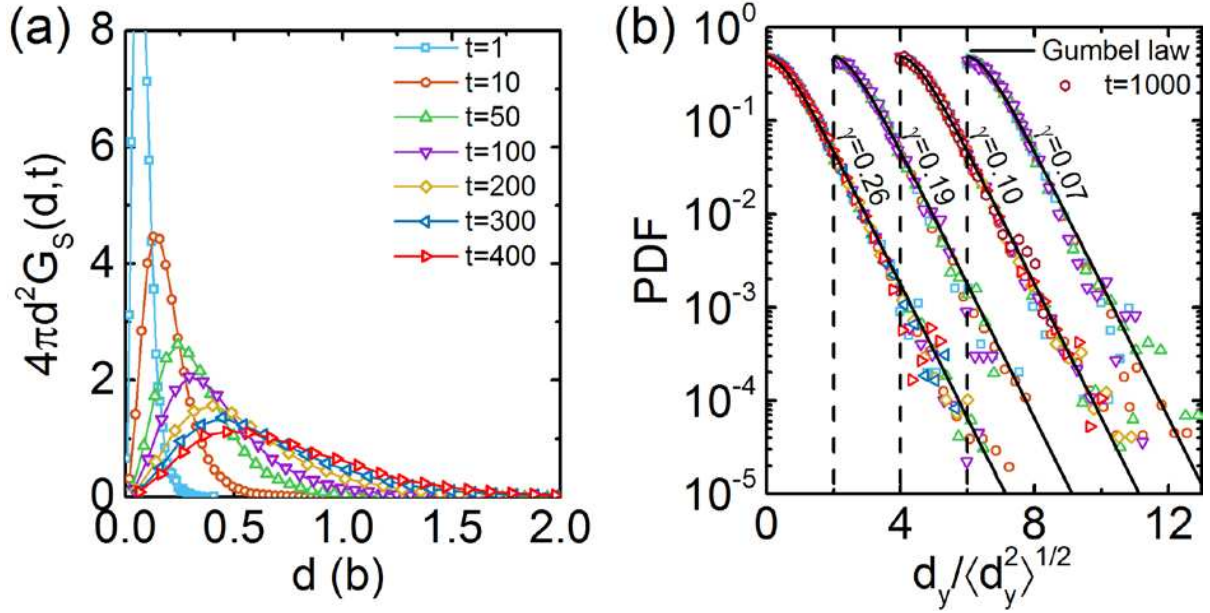

**Figure 2| Time evolution of the self-part of the van Hove function and probability density function of the translational displacement along the *y* direction.**

a) The van Hove function $G_s(d,t)$ at different times $t$ for $\gamma =0.26$. The times are given in the legend. b) Probability density functions of particle displacements in the *y*-direction, i.e. the shear direction (Fig. 1a), as function of $d_y/\langle d_y^2 \rangle^{1/2}$ for different values of $\gamma$; symbols are the same as in panel a). (For the sake of clarity the data for $\gamma =0.19$, 0.10, and 0.07 has been shifted horizontally by 2.0, 4.0, and 6.0, respectively.) Also included is a fit to the data with a Gumbel law (solid line)

$$f(d_y) = A(\lambda)\exp\left(-\frac{|d_y|}{\lambda} - \exp\left(-\frac{|d_y|}{\lambda}\right)\right)$$ with $\lambda =0.605$. Here $A(\lambda)$ is the normalization constant and its value is 1.313. The same quantitative behavior is found in the *x* and *z* directions.

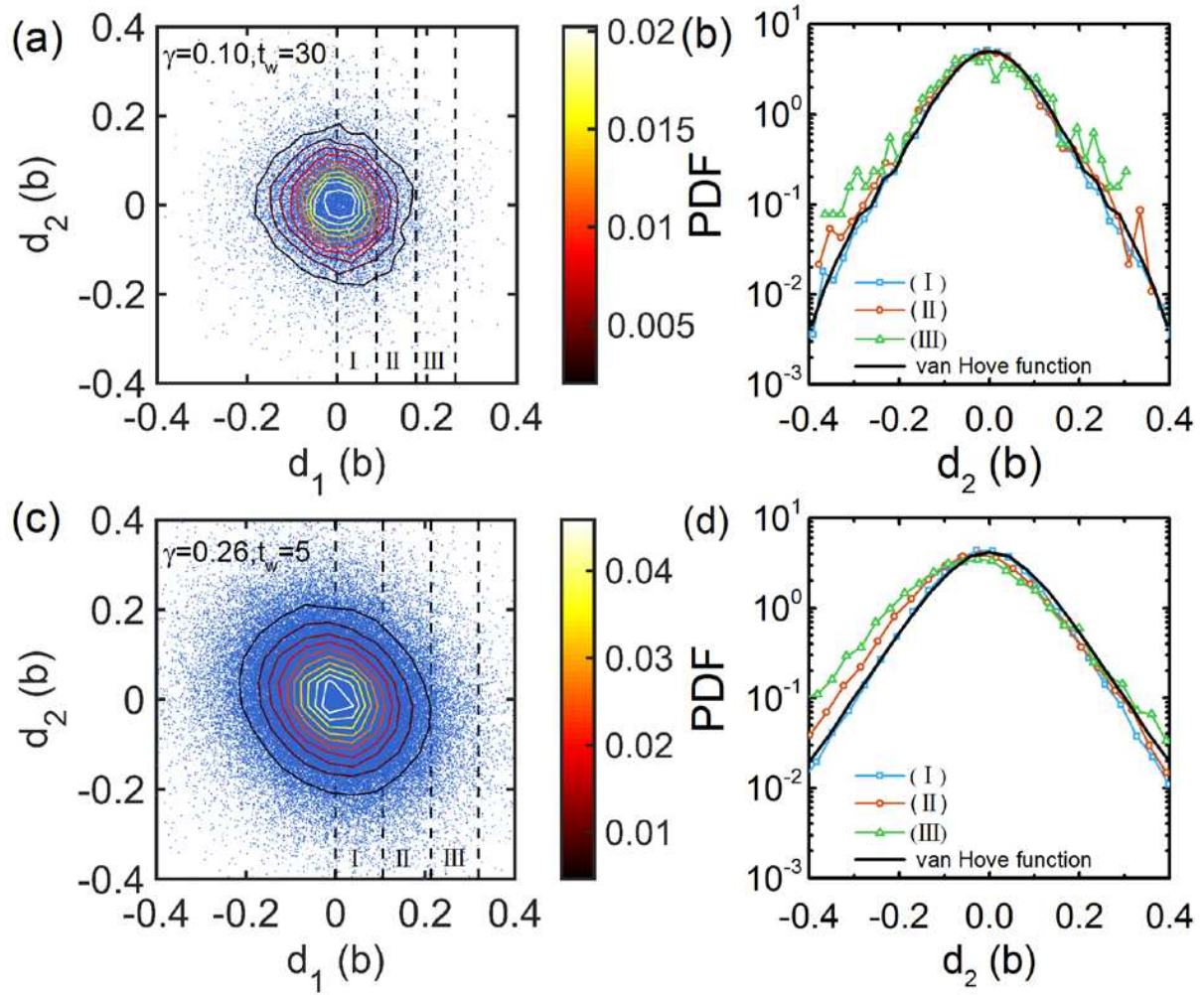

**Figure 3| Scatter plots of the displacements for two consecutive time lags and the conditional probability density function of $d_2$ to demonstrate the presence of memory effect at large $\gamma$.**

a) Scatter plot of $d_2$, the displacement in the time interval [30, 60], as function of $d_1$, the displacement in the time interval [0, 30], $\gamma =0.10$. The vertical dashed lines at $d_1 =0$, $\langle d_1^2 \rangle^{1/2}$, $2\langle d_1^2 \rangle^{1/2}$, and $3\langle d_1^2 \rangle^{1/2}$ are used to define the regions I, II, and III respectively. b) The conditional PDF of $d_2$ when $d_1$ belongs to region I, II, III in a). The solid line is the full PDF of $d_2$, i.e. the van Hove function in y direction. c) Scatter plot of the displacement $d_2$ in the time interval [5, 10] as function of $d_1$, the displacement in the time interval [0, 5], $\gamma =0.26$. The time intervals in a) and c) are chosen so that the corresponding TMSDs are comparable (see Fig. 1b). The vertical dashed lines correspond to $d_1 = 0$, $\langle d_1^2 \rangle^{1/2}$, $2\langle d_1^2 \rangle^{1/2}$ and $3\langle d_1^2 \rangle^{1/2}$. The colour scale (shown as contours) in (a) and (c) corresponds to the density of points. d) The conditional PDF of $d_2$ when $d_1$ belongs to region I, II, III in c). The solid line is the full PDF of $d_2$.

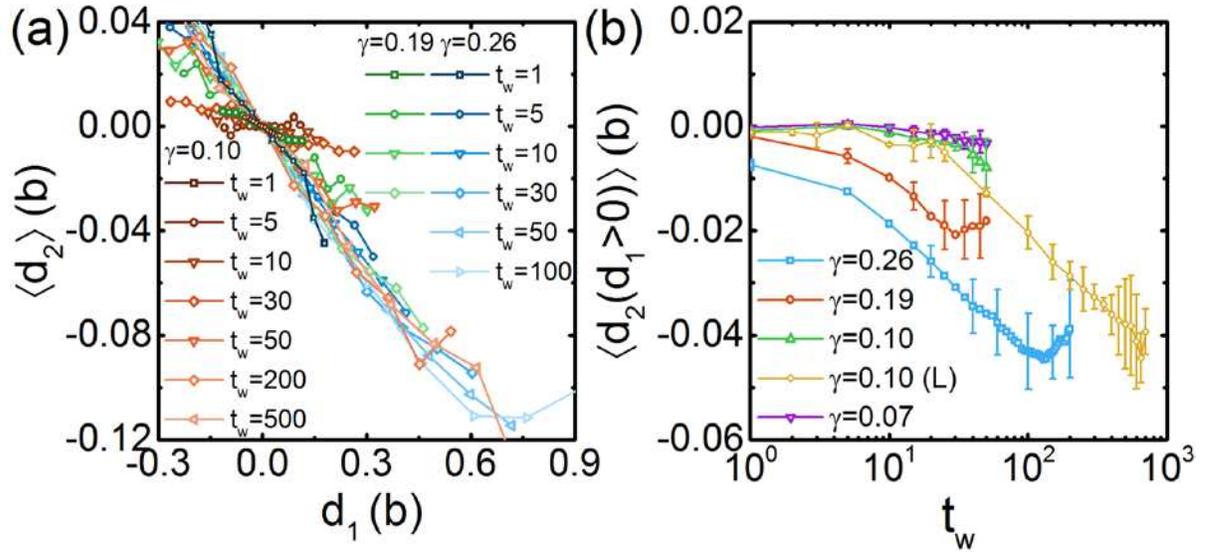

**Figure 4| Memory effect as a function of particle displacement and time.**

a) Average of $d_2$ as a function of $d_1$ for different times and $\gamma$. For fixed time the averages increase (in absolute value) with $\gamma$ showing that large strain amplitude gives rise to stronger memory effects. The data for $\gamma = 0.10$ shows that even for small strain amplitude the memory effect is seen, although at times that are longer than with larger $\gamma$. For displacements larger than approx. 0.6b, the curves start to flatten (see $\gamma = 0.26$ for $t_w = 100$), indicating that once a particle has been displaced to about this distance it starts to lose its memory. Qualitatively the same behavior is found in the $x$ and $z$ directions. b) The $t_w$-dependence average of $d_2$ for $d_1 > 0$ showing that the memory effect increases with time before it starts to decreases again at long times. Error bars are calculated as the standard deviation of $d_2$ in the three directions.

# Supplementary Information
# Granular materials flow like complex fluids

Binquan Kou, Yixin Cao, Jindong Li, Chengjie Xia, Zhifeng Li, Haipeng Dong,
Ang Zhang, Jie Zhang, Walter Kob*, and Yujie Wang*

# Contents

## 1. Supplementary Discussion

(1) Experimental setup

(2) Preparation of the sample

(3) Structural properties of the sample

    A) Effect of the walls on the structure

    B) Evolution of the volume fraction Φ under shear

    C) The static structure factor $S(q)$.

(4) Determination of the rotational dynamics

(5) Intermediate scattering function

(6) Origin of the Gumbel law

(7) Comparison of Gumbel and q-Gaussian fitting.

## 2. Supplementary Movies

## 3. Supplementary References

# 1) Supplementary Discussion

**(1) Experimental setup**

The rectangular shear cell (see Fig. 1a in the main text for a schematic diagram) is made of acrylic plates and has a size of 40.2b×43b×22.6b (25.5cm×27.25cm×14.3cm). The front and back plates are permanently fixed on the baseplate. A plate is laid on top of the particle packing, and is constrained by linear guides on the front and back plates, with the result that it can undergo only motion in the vertical direction. The total weight of the particles is 8.3kg and the weight of the top plate is 16kg and thus provides a constant normal pressure of 2.3kPa. This extra weight from the top plate helps to minimize the relative pressure gradient in the packing along the gravity direction. Using a materials testing machine (Zwick//Roell Z100, Germany) we have determined Young's modulus $E$ of the particles and found it to be $4.13 \pm 0.18$ GPa, a value that is close to the one found in the literature for PVC[31]. From this value and the applied pressure one can estimate that the deformation of the particles is given by $d = \left( \frac{9F^2}{16(b/2)E^2} \right)^{1/3} \approx 1.778 \mu m \approx 2.81 \times 10^{-4} b$ [32]. Here $F$ is the force on a bead that we estimate to be given by the total weight ($= 24.3 \text{kg} \times 9.8 \text{m/s}^2 = 238.14$N) divided by the total number of beads in the bottom layer (=324) which gives $F$=0.735N. Hence we can conclude that under our experimental conditions the particles keep their shape and hence can be considered as hard ellipsoids[22].

The side plates are linked by hinges to the bottom plate which in turn is attached to a linear motor stage, whose horizontal displacement generates the shear on the shear cell. The shear direction is defined as the *y* direction and the vertical direction as *z* direction. During a shear cycle, the cell will first be sheared in one direction to the designated shear strain; then the shear is reversed and the cell will be sheared in the other direction to reach the same negative strain.

Finally, the whole cell is returned to its origin geometry to complete the cycle.

**(2) Preparation of the sample**

Before we started the measurements with the CT scanner to obtain the positions and orientations of the particles we have cycled the system for a long time in order to allow it to reach a steady state. To estimate the time it takes the system to reach a steady state we have monitored the height of the movable top plate as a function of time since this type of measurement allows to obtain in a simple manner a reliable estimate of the total volume occupied by the particles. We have found that the average height decreases quickly during the first hundred cycles and then shows a very slow and weak decrease. Since after these first hundred cycles the packing fraction in the central part of the system, i.e. in the region which we have used for analyzing our CT measurements (see next section) is not changing anymore with time, see Fig. SI2, we can conclude that the central part of the system has indeed reached the steady state, i.e. the changes in its structure are sufficiently small that they can be neglected. The number of cycles for the preparation and the number of cycles used for the measurements is given in Tab. SI1.

| Shear strain ($y$-direction) | Number of cycles to prepare initial state | Number of cycles for measurements |
|---|---|---|
| 0.26 | 300 | 615 |
| 0.19 | 550 | 125 |
| 0.10 | 1500 | 125 |
| 0.10 (long run) | 1500 | 1850 |
| 0.07 | 2400 | 125 |

Tab.SI1 Experimental protocol used to prepare the system and to make the measurements of its properties.

### (3) Structural properties of the sample

A) Effect of the walls on the structure

In Fig. SI1 we plot the normalized local particle number density $\rho_i/\langle\rho_i\rangle$, $i=x, y, z$ of the system. Clear layering effects are found near the boundary region along all three axes. To minimize the potential influence of these layering effects, we excluded particles whose distances to the boundary are smaller than 5b. However, as shown in Fig. SI1c, a weak layering effect still exists along the $z$ direction in the center region. To estimate the influence of this small layering effect, we have divided the central part of the box along the $z$ direction into three parts of equal height. We have carried out the same analyses regarding the relaxation dynamics described in the main text for each of these three regimes and found no quantitative differences among them. Therefore we can conclude that our results are not affected by wall effects.

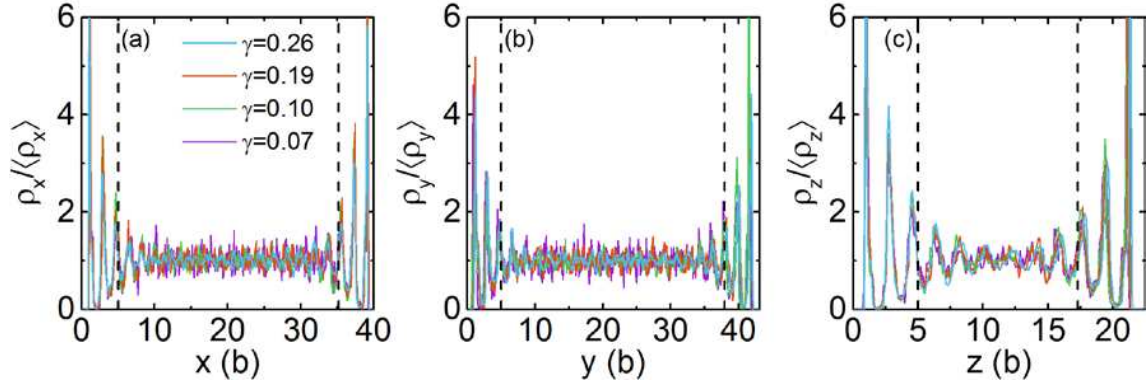

Fig.SI1. The normalized local number density $\rho_i/\langle\rho_i\rangle$, $i=x, y, z$, in the system as a function of the three coordinates. The vertical dashed lines denote the boundaries of the central region we have chosen for the subsequent analysis.

B) Evolution of the volume fraction Φ under shear

The volume fraction $\Phi$ is determined by $\Phi = \sum_j V_j \Big/ \sum_j V_{cell}$, where $V_j$ and $V_{j,cell}$ are the volume of particle $j$ and its Voronoi volume, respectively, and the sums are taken over all particles that are in the central volume. The time dependence of $\Phi$ is shown in Fig. SI2. The quite constant values of $\Phi$ during measurements indicate that the system has indeed reached the steady state for all $\gamma$ investigated. Also note that for the second run for $\gamma=0.10$ we find a volume fraction that is compatible with the one we found for first run for this strain amplitude. This indicates that our results are reproducible and do not show large sample to sample fluctuations.

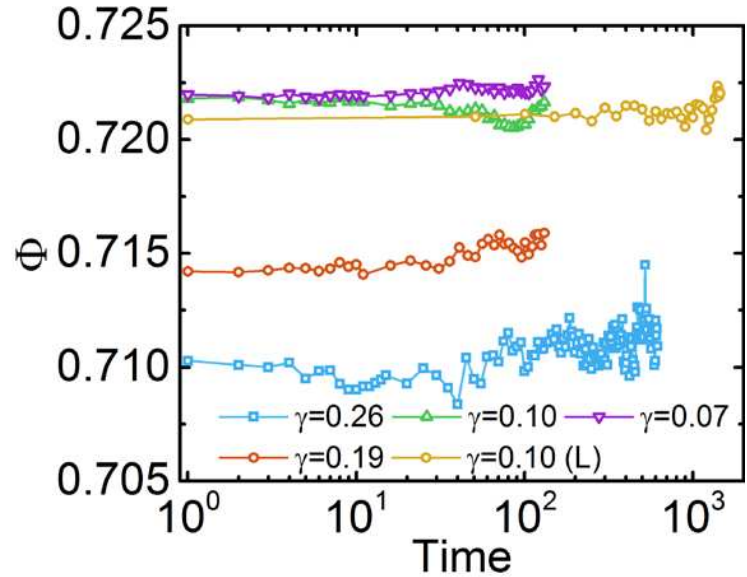

Fig. SI2. Evolutions of the volume fraction $\Phi$ of the system during the cyclic shear measurements at different shear strain $\gamma$. Note that for $\gamma=0.10$ we have two curves that stem from two completely independent measurements. The fact that the corresponding packing fractions are compatible shows that sample to sample fluctuations are rather small.

C) The static structure factor $S(q)$.

We have calculated the static structure factor $S(q)$ from the particle position via

$$S(q) = \frac{1}{N}\sum_{k=1}^{N}\sum_{k\neq j}^{N}\left\langle \exp\left[\,i\mathbf{q}\cdot(\mathbf{r}_k-\mathbf{r}_j)\right]\right\rangle. \tag{1}$$

As shown in Fig. SI3, we find that for all values of $\gamma$, $S(q)$ shows the typical form found in disordered systems[5] thus indicating that there is no sign of crystallization. (The peak at very small $q$ is related to the boundary effect of the window function used for the Fourier-transform.)

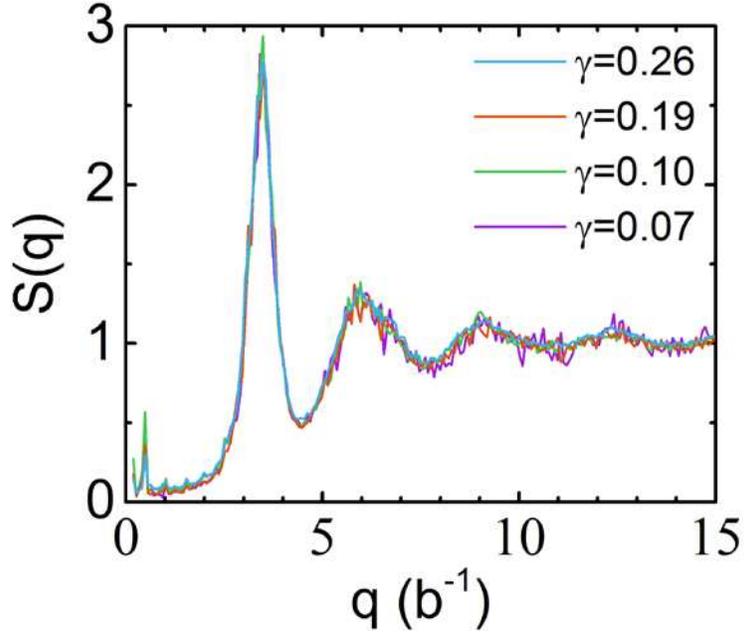

Fig. SI3. The static structure factor $S(q)$ of the system in steady state prepared for different shear strain $\gamma$.

**(4) Determination of the rotational dynamics**

In order to quantify the rotational motion of the particles, we have calculated the time integral of the angular increment of each cycle, $\theta_j(t) = \int_0^t \omega_j(t')dt'$, where the modulus and direction of $\omega_j(t)$ are given by $\cos^{-1}\left[e_j(t)\cdot e_j(t+1)\right]$ and the vector $e_j(t)\times e_j(t+1)$, respectively.

From this we can then define an unbounded rotational mean squared displacement (RMSD) as $\langle |\boldsymbol{\theta}(t)|^2 \rangle$ which is shown in Fig. 1c of the main text.

**(5) Intermediate scattering function**

From Fig. 1b of the main text we concluded that this system does not show a cage effect for any value of $\gamma$. To test whether this conclusion is indeed correct we have calculated from the particle positions $\mathbf{r}_j$ the self-intermediate scattering function, $F_s(q,t)$, i.e. the density correlation function for wave-vector $q$[5, 33]:

$$F_s(q,t) = \frac{1}{N}\left\langle \sum_{j=1}^{N} \exp\left[-i\mathbf{q}\cdot(\mathbf{r}_j(t) - \mathbf{r}_j(0))\right]\right\rangle \quad (2)$$

The time dependence of $F_s(q,t)$ is shown in Fig. SI4 for two wave-vectors: In panel a) for $q = 3.49b^{-1}$, i.e. the position of the main peak in the static structure factor $S(q)$, and in panel b) for $q = 4.46b^{-1}$, i.e. the location of the first minimum in $S(q)$ (see Fig. SI3). We recall that for thermal glass-forming systems the cage effect makes that $F_s(q,t)$ shows a two-step relaxation process, i.e. at intermediate times it shows a plateau before it decays at long times to zero[5]. The curves in Fig. SI4 show that for none of the γ's there is any indication for the presence of such a plateau, demonstrating thus that for this system the cage effect is completely absent. We have also calculated the collective intermediate scattering function $F_s(q,t)$ [5,33] and found the same qualitative behavior.

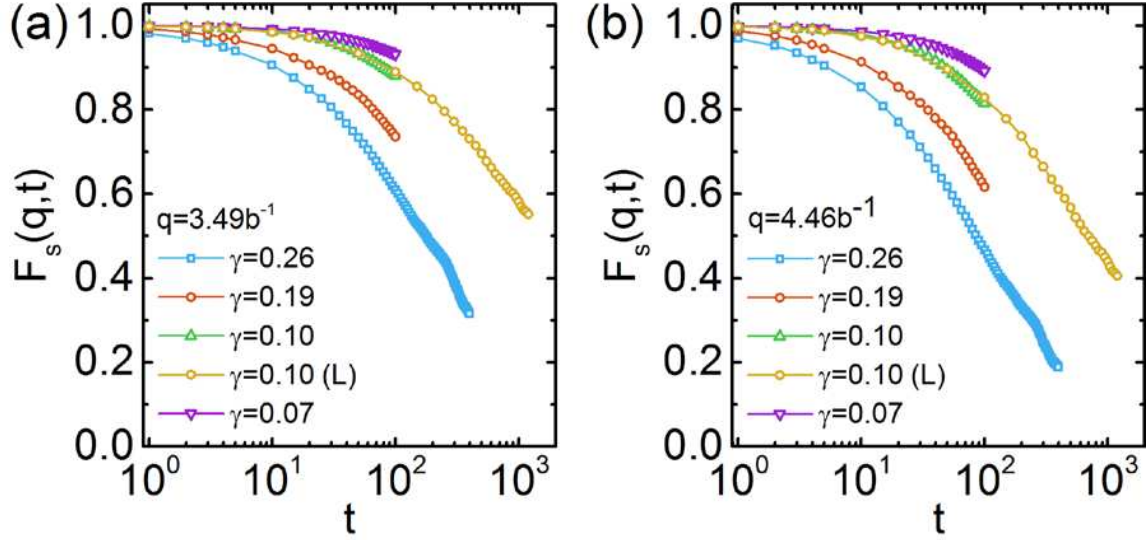

Fig. SI4: Time dependence of the self-intermediate scattering function $F_s(q,t)$ for different strain amplitudes. Panels a) and b) are for $q = 3.49\text{b}^{-1}$ and $q = 4.46\text{b}^{-1}$, respectively. Note that the correlators decay without any sign of a two-step relaxation, thus showing that in this system the particles are not caged.

**(6) Origin of the Gumbel law**

Here we outline the arguments why the PDF for the displacements of the particles is described well by the Gumbel law, as long as the displacement of the particles is not too large, i.e. the particles have not yet changed a significant fraction of their nearest neighbors (see Fig. 2b of the main text). For a granular system with no friction every mechanically stable configuration will correspond to an arrangement of the particles in which the total gravitational energy of the system is in a local minimum. Although this is similar to the situation found in a glass at *finite* temperature (i.e. a thermal system), there is one important difference: In thermal systems every particle will vibrate around its equilibrium position with an amplitude that is *finite* and that is proportional to the square root of the temperature. Thus an arrangement of particles in which such a vibrational motion would allow anyone of the particles to overcome the local barrier (saddle point) in the potential energy is unstable. In other words, every particle in the thermal system must have a typical distance from its closest local saddle point that is finite and not too small. In contrast to this, a granular system at rest is at zero temperature and hence there is no need that a particle has a minimum distance from is nearest saddle point. As a consequence, even the slightest perturbation of a granular system will make that some of the particles become destabilized, in contrast to thermal systems. The presence of friction does not alter this conclusion significantly.

The cyclic shear makes that the particles explore their neighborhood and if they encounter a mechanical instability they will relax to a new local minimum. Note that the details on how this exploration occurs will depend on the properties of the particles in that, e.g., for perfectly smooth particles (with or without dynamic friction) the exploration is completely deterministic, thus allowing for the existence of limit cycles[17]. In contrast to this, one can expect that particles that have some roughness will show an exploration of their neighborhood that is similar to a random walk, since the roughness will induce a weak stochastic component to their motion. Also the static friction will induce random noise into the trajectories of the particles since every time that the system arrives at the maximal strain of the oscillation, the motion stops and the presence of a static friction will make that the return path of a particle will not necessarily coincide with the trajectory that leads to the turning point. Since our particles do indeed have some roughness and static friction, we thus can expect that for weak $\gamma$ their motion is diffusive, with diffusion constant $D$, and hence the TMSD shows a linear time dependence, in agreement with what we find (see Fig. 1b of the main text) and a Gaussian shape of the PDF for small distances (Fig. 2b of the main text). Thus this Gaussian distribution can be written as

$$P_g(d,t) = \frac{1}{\sqrt{2\pi\Delta(t)^2}} \exp\left(-\frac{d^2}{2\Delta(t)^2}\right) \qquad (3)$$

where $\Delta(t) = \langle d^2 \rangle^{1/2} = (2Dt)^{1/2}$ is the square root of the mean squared displacement at time $t$. (Note that for the sake of simplifying the notation we consider here just a one dimensional motion characterized by a displacement $d$.)

This random walk of the particles makes that after some time certain particles will reach a saddle point in the energy landscape and thus they will relax, i.e. flip over or fall down into a "hole", thus making an irreversible relaxation event (IRE). We define $\varepsilon$ as the distance that a given particle has to move from its starting position to reach this saddle point. We have argued above that for a granular system a particle in a stable packing can be arbitrarily close to a saddle point, which implies that the probability distribution of this distance, $W(\varepsilon)$, is not small even at small $\varepsilon$. In a first approximation we can assume that $W(\varepsilon)$ is basically a constant for $\varepsilon \leq \sigma$ and becomes zero for $\varepsilon$ on the order of $\sigma$, the size of the particles, i.e. there are no cavities that are larger than the size of the particles. Because of the normalization of $W(\varepsilon)$ we thus can approximate it by $1/\sigma$.

Due to the diffusive motion of the particles, $\varepsilon(t)$ is a random variable that makes the same type of diffusive dynamics as the position of the particles, i.e. $[\varepsilon(t) - \varepsilon(0)]^2 = 2Dt$. Destabilization of the particle means that $\varepsilon(t)$ has become zero, i.e. after a time $t$ a particle that had a $\varepsilon(0) \leq (2Dt)^{1/2}$ is likely to have undergone an IRE. Let us call $h_0$ the typical distance that a particle moves because of an IRE. ($h_0$ is certainly less than $\sigma$ because we assume that there are no cavities and a typical value will be around $\sigma/2$.) Let us assume that the probability that a particle has undergone an IRE is small, i.e. $(2Dt)^{1/2}/\sigma = p_0 \ll 1$. The probability that a particle has made in the time $t$ a number $k$ of such events $(k = 1, 2, 3, ...)$ is thus proportional to $p_0^k \approx \exp(-k/p_0)$ and in doing so it has moved $d = O(h_0 k)$. Thus we have

$$P_{IRE}(d,t) \approx \exp(-k/p_0(t)) \approx \exp\left(-\frac{d\sigma}{h_0(2Dt)^{1/2}}\right) = \exp(-\frac{d\sigma}{h_0\Delta(t)}). \qquad (4)$$

The probability that a particle has in time $t$ moved by a distance $d$ is thus given by the sum of these two processes:

$$P_T(d,t) = P_g(d,t) + P_{IRE}(d,t) \approx \exp\left(-\frac{d^2}{2\Delta(t)^2}\right) + \exp(-\frac{d\sigma}{h_0\Delta(t)}). \qquad (5)$$

(For the sake of simplicity we assume here that the two processes have the same weight.) Since $\sigma/h_0 \geq 1$, the first term on the right hand side will dominate the second one if $d$ is small, i.e. for displacements that are small with respect to $\Delta(t)$, the PDF will basically be a Gaussian. If $d \approx \Delta(t)$ the two terms on the right hand side become comparable. Thus a plot of $P_T$ as a function of $d/\Delta(t)$ will show at $d/\Delta \approx h_0/\sigma = O(1)$ a crossover from the Gaussian behavior to the exponential dependence, in agreement with our finding (see Fig. 2b of the main text).

We now compare the slopes of the two distributions at $d = \Delta$ (or rather of the logarithm of the distributions). One finds immediately

$$\frac{d\log[P_g(d)]}{dd} = \frac{-d}{\Delta^2} \quad \text{and} \quad \frac{d\log[P_{IRE}(d)]}{dd} = -\frac{\sigma}{h_0\Delta}. \qquad (6)$$

At the crossing point $d = \Delta$ the two distributions have thus the same slope $-1/\Delta$, since $\sigma/h_0 \approx 1$ which means that there is *no* kink in the total distribution. As a consequence the total PDF can be approximated well by a Gumbel law since this functional form has also a Gaussian dependence at short distances and an exponential tail at large distances. Equation (4) shows that the parameter $\lambda$ of the Gumbel distribution, see caption of Fig. 2 in the main text, is thus given by $h_0 \cdot \Delta(t)/\sigma$, i.e. it has a simple geometrical interpretation. We point out that other functional forms can certainly be used to describe these two limiting behaviors, such as the q-Gaussian discussed below. However, because of its simplicity and the fact that it has only one free parameter we will here give preference to the Gumbel law.

Note that this result holds for all values of time, i.e. once the PDF has been normalized by $\Delta(t)$ its shape is independent of time, in agreement with our results shown in Fig. 2b of the main text. Furthermore we recall that the shape of the PDF is independent of $\gamma$ (Fig. 2b). This result can be rationalized by noticing that our argument for the observation of the Gumbel law depends on the presence of a diffusive dynamics on short length scales which is independent of $\gamma$, if the distribution is normalized by $\Delta(t)$. Thus even if this dynamics is not *perfectly* diffusive, because, e.g. the presence of a memory effect, the final result will not be changed significantly, since, for symmetry reasons, the central part of the PDF can always be approximated well by a parabola, i.e. a function that for small arguments is similar to a Gaussian.

Finally we point out that the existence of the Gumbel law is directly connected to the fact that the probability that a given particle makes an IRE is (statistically) related to its displacement since the last time it made such a jump which in turn is proportional to the square root of the TMSD. This is in contrast to the case of the thermal systems in which the time scale at which a hopping occurs is independent of time and hence no Gumbel law is observed[34].

**(7) Comparison of Gumbel and q-Gaussian fitting.**

In previous studies it has been found that $G_s(d,t)$ can be fitted well with a so-called q-Gaussian with a parameter $q$ that depends on $\gamma$. This q-Gaussian distribution is given by[35,36]:

$$G_q(d) = \frac{\sqrt{(q-1)}\,\Gamma\!\left[\dfrac{1}{(q-1)}\right]}{\sqrt{\pi A}\cdot\Gamma\!\left[\dfrac{(3-q)}{2(q-1)}\right]} \left[1\;(1\text{-}q)\dfrac{d^2}{A}\right]^{1/(1-q)} \tag{7}$$

where $A$ and $q$ are fit parameters and $\Gamma$ is the Gamma-function. We have fitted our data with this functional form as well and found that the resulting fit does not give a very satisfactory fit of the data in the tails (see Fig. SI5a). However, close to the peak the q-Gaussian does give a better description of the data (see Fig. SI5b). In Figs. SI5c and *d* we show the difference between the data and the two fits with the two functional forms. Although on overall the q-Gaussian does give a better description of the data, one has to recall that this functional form has two fit parameters whereas the Gumbel distribution has only one. In addition we note that the choice of the Gumbel distribution can be rationalized by the arguments given in the preceding section whereas for granular materials the q-Gaussian law does not really have a solid foundation. In view of these problems it is therefore difficult to tell which functional form, if any, is more appropriate. Although the difference between our results and the ones of Refs. 11 and 15 might be related to the fact that we have a three dimensional systems whereas previous investigations considered two dimensional ones, the mechanism leading to the Gumbel distribution is quite general and hence might be present in 2D systems as well.

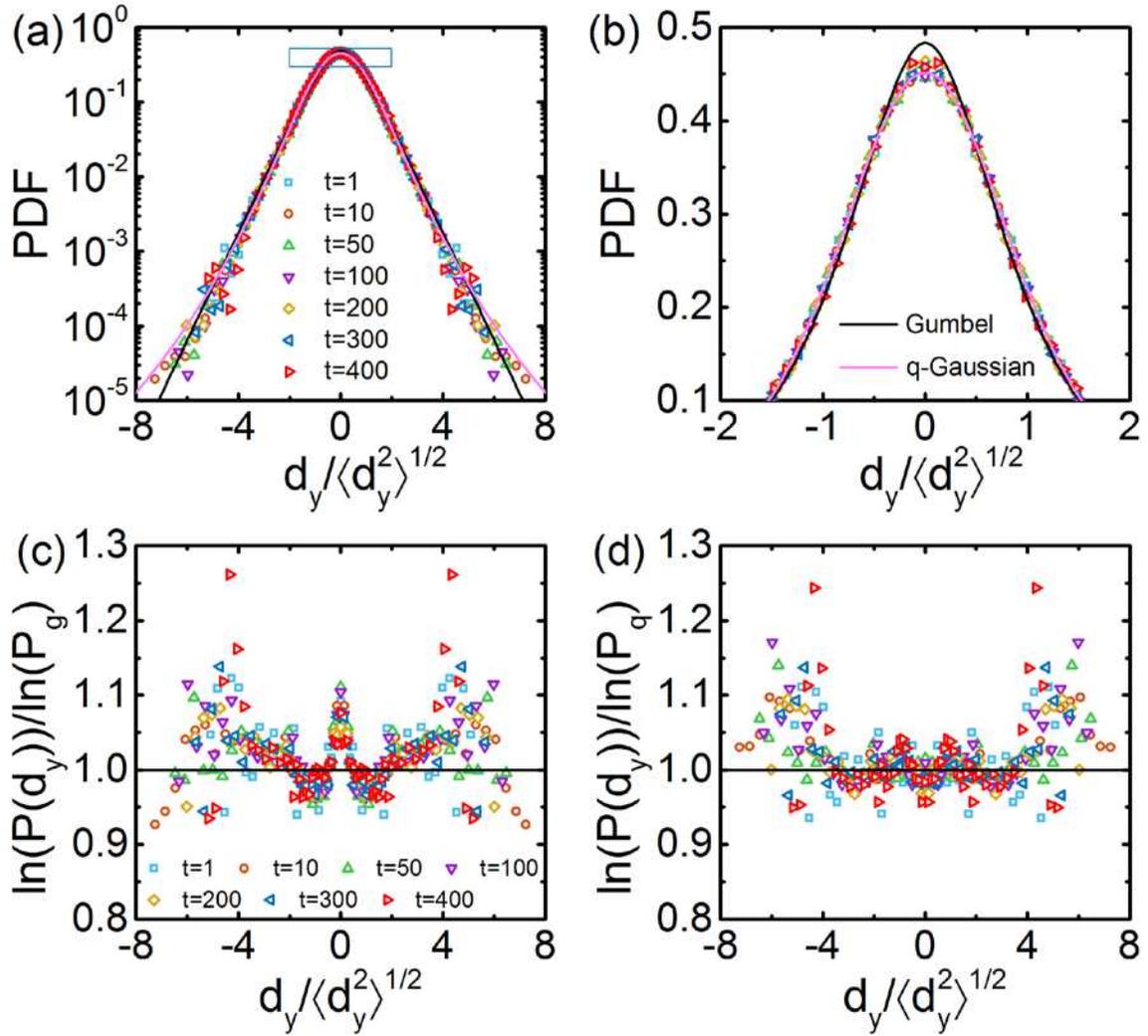

Fig. SI5. PDF in *y*-direction for $\gamma=0.26$. Comparison of the fits with the Gumbel distribution (black line) and the q-Gaussian distribution (pink line) for the whole accessible range in the displacement (panel a). Panel b) shows an enlarged view of the blue rectangle in (a). Panels c) and d) show, respectively, the ratio between the logarithm of the PDF between the data and the Gumbel distribution and the data and the q-Gaussian distribution.

## 2) Supplementary Movies

**Supplementary Movies: The dynamics of the particles during the cycling experiment.**

The two movies M1 and M2 show a view of the system from the top and the side, respectively. M1 shows a horizontal cut through the middle of the sample and only the particles in the lower half of the central region of the box are displayed. For M2 we made a vertical cut through the middle of the sample and show only the particles in the sector with large $x$. The value of $\gamma$ is 0.26 and the movie covers the 615 cycles of the measurement with the CT scanner. Note that on the time scale considered, most of the particles move less than their long axis (see Fig. 1b of the main text). Also note that the system shows no evident signature for convective motion.

## 3) Supplementary Reference

5    Binder, K. & Kob, W. *Glassy materials and disordered solids: An introduction to their statistical mechanics*. (World Scientific, 2011).

11   Ren, J., Dijksman, J. A. & Behringer, R. P. Reynolds pressure and relaxation in a sheared granular system. *Phys. Rev. Lett.* **110**, 018302 (2013).

15   Radjai, F. & Roux, S. Turbulentlike fluctuations in quasistatic flow of granular media. *Phys. Rev. Lett.* **89**, 064302 (2002).

17   Royer, J. R. & Chaikin, P. M. Precisely cyclic sand: self-organization of periodically sheared frictional grains. *Proc. Natl Acad. Sci. USA* **112**, 49-53 (2015).

22   Singh, A., Magnanimo, V., Saitoh, K. & Luding, S. The role of gravity or pressure and contact stiffness in granular rheology. New J. Phys. 17, 043028 (2015).

31   Titow, M. *PVC Technology*. (Springer Science & Business Media, Berlin,1984).

32   Johnson, K. L. & Johnson, K. L. *Contact mechanics*. (Cambridge University Press, Cambridge, 1987).

33   Hansen, J.P. & McDonald, I. R. *Theory of simple liquids*. (Elsevier, Amsterdam, 1990).

34   Chaudhuri, P., Berthier, L. & Kob, W. Universal nature of particle displacements close to

| | |
|---|---|
| | glass and jamming transitions. *Phys. Rev. Lett.* **99**, 060604 (2007). |
| 35 | Umarov, S., Tsallis, C. & Steinberg, S. On a q-Central Limit Theorem Consistent with Nonextensive Statistical Mechanics. *Milan J Math* **76**, 307-328 (2008). |
| 36 | Picoli, S., Mendes, R. S., Malacarne, L. C. & Santos, R. P. B. q-distributions in complex systems: a brief review. *Braz. J. Phys.* **39**, 468-474 (2009) |